\documentclass[a4paper,12pt]{article}
\usepackage{setspace}
\usepackage[english]{babel}
\usepackage{authblk}
\usepackage{palatino}
\usepackage{svg}

\usepackage[utf8]{inputenc}
\DeclareUnicodeCharacter{2009}{\,}
\DeclareUnicodeCharacter{03BC}{$\mu$}
\DeclareUnicodeCharacter{03B1}{$\alpha$}
\DeclareUnicodeCharacter{2212}{-}

\usepackage[a4paper,top=2cm,bottom=2cm,left=3cm,right=3cm,marginparwidth=1.75cm]{geometry}

\usepackage{amsmath}
\usepackage{graphicx}
\usepackage[colorlinks=true, allcolors=blue]{hyperref}

\title{Femtosecond all-optical coherent control of spin polarization in altermagnets}
\author[1]{K. J\"ackel}
\author[1]{H. Grisk}
\author[1]{N. Dornquast}

\affil[1]{Institut für Physik, Universität Greifswald, Felix-Hausdorff-Straße 6, 17489 Greifswald, Germany}

\author[2]{M. Gaerner}

\author[2]{G. Reiss}
\affil[2]{Faculty of Physics, Bielefeld University, Universit\"atsstra\ss e 25, 33615 Bielefeld, Germany}

\author[2,3]{T. Kuschel}
\author[1]{J. Walowski}
\author[1]{M. Münzenberg}

\affil[3]{Institute of Physics, Johannes Gutenberg University Mainz, Staudinger Weg 7, 55128 Mainz, Germany}

\begin{document}
\maketitle

\begin{abstract}

 Altermagnets constitute an emerging materials platform for spintronic technologies by combining compensated magnetic order with ferromagnet-like spin-split electronic bands. Here, we investigate the proposed d-wave altermagnetic material RuO$_2$ using circularly polarized ultrashort laser pulses. Time-resolved magneto-optical Kerr effect measurements, which are intrinsically sensitive to surface and interface states, reveal the ultrafast spin response of RuO$_2$. In contrast to the demagnetization dynamics characteristic of conventional ferromagnets, we observe a distinct coherent contribution to the complex Kerr rotation that appears during the light–matter interaction and lasts for $\sim200\,\mathrm{fs}$. Similar signatures have been associated with spin–momentum locking and directional band splitting in spin-split surface states of topological insulators as well as spin–orbit–coupled semiconductors and are governed by a finite Raman coherence time. We interpret this coherent response as evidence for transient spin-polarized surface states in RuO$_2$, consistent with the emergence of altermagnetic surface states that are directly relevant to spin-polarized transport at surfaces and interfaces.

\end{abstract}

\section{Introduction}

\begin{figure}[!ht]
    \includegraphics[width=0.9\linewidth]{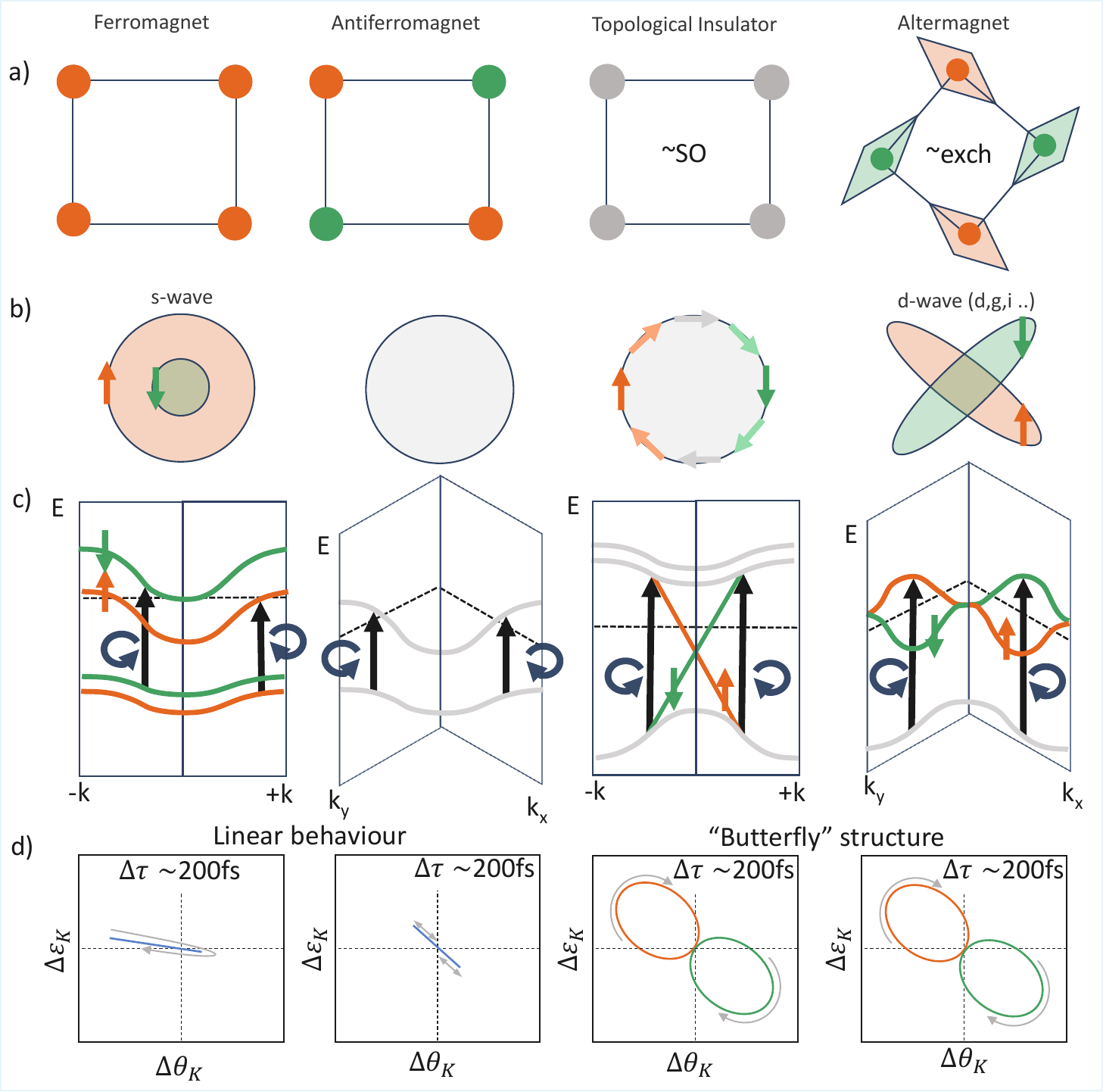}
    \caption{a) Illustrates the most basic crystal structures with either up- or down-oriented, or zero magnetic moment, in orange, green, and gray, respectively, for ferromagnets, antiferromagnets, topological insulators, and altermagnets (from left to right). b) A cross-section through the Fermi surfaces for all four material systems. c) We apply selective pumping processes by utilizing circularly polarized photons to enhance specific electron transitions. d) This leads to material-distinct magnetization dynamics, measured by time-resolved magneto-optical Kerr-effect, apparent in the relation between the real $\Delta\theta_K$ and the imaginary $\Delta\epsilon_K$ parts of the complex Kerr rotation. Transient spin polarized states (e.g. topological insulators and altermagnets) produce characteristic "butterfly" structures in materials with $\vec{k}$ dependent electron spin within the first $200\,\mathrm{fs}$ after the excitation with ultrashort laser pulses $\sim 45\,\mathrm{fs}$. In (anti-) ferromagnets, the electron spin is, in first order, independent of $\vec{k}$, revealing a linear relation.}
    \label{fig:MaterialsOverview}
\end{figure}

The integration of spintronics into modern information technology has catalyzed an intensive search for emergent physical effects and materials that can be seamlessly integrated into scalable device platforms. Using spin-dependent transport and collective phenomena, spintronic architectures promise reduced power consumption, increased functional density, and compatibility with existing semiconductor technologies. This positions spin-based computing as a compelling route toward energy-efficient, on-chip implementations of neuromorphic, probabilistic, and hybrid quantum information processing \cite{oberbauer_magnetic_2025}.

Altermagnets have emerged as a new magnetic paradigm. The altermagnetic spin order combines anisotropic, non-relativistic, spin-split bands with a compensated, collinear spin arrangement \cite{smejkal_crystal_2020, smejkal_emerging_2022, smejkal_beyond_2022, jungwirth_altermagnetism_2025}. This occurrence expands the material base for spin-based information technologies by combining a compensated collinear spin structure with spin-split electronic bands, thereby overcoming the long-standing limitations of conventional collinear antiferromagnets in device operation and readout. The symmetry-driven origin of this effect enables efficient electrical control and detection of spin-polarized currents, while preserving the intrinsic robustness of the compensated magnetic order. Consequently, altermagnetic materials offer an attractive route to scalable spintronic architectures that enable high-efficiency magnetoresistive effects and current-induced torques, with direct relevance to memory, logic, and neuromorphic computing \cite{jungwirth_altermagnetic_2025}.

Generally, materials with an antiferromagnetic spin arrangement exhibit spin-degenerate bands because of a preserved PT-symmetry, prohibiting large spin-polarizations arising from spin-split bands. In altermagnets, however, the opposite spin sublattices are connected by rotations, such as $\mathrm{C}_4$ and $\mathrm{C}_6$, instead of translation or inversion \cite{smejkal_emerging_2022}. This leads to momentum-dependent spin polarization, producing alternating spin textures in k-space. They are even in momentum s(k)=s($-$k), and are classified by orbital symmetry representations, such as d-wave (e.g. $\mathrm{RuO_2}$, $\mathrm{Mn_5Si_3}$ \cite{smejkal_emerging_2022, reichlova_observation_2024, badura_observation_2025, han_electrical_2024}, $\mathrm{MnF_2}$ \cite{smejkal_emerging_2022, bhowal_ferroically_2024,faure_altermagnetism_2025}) or g-wave (e.g. MnTe \cite{kriegner_magnetic_2017,gonzalez_betancourt_spontaneous_2023,mazin_altermagnetism_2023,krempasky_altermagnetic_2024, amin_nanoscale_2024, osumi_observation_2024}, CrSb \cite{reimers_direct_2024, ding_large_2024} or $\mathrm{Fe_2O_3}$ \cite{hoyer_altermagnetic_2025, galindezruales_revealing_2025}), and higher orders \cite{smejkal_emerging_2022}, giving altermagnets a unique fingerprint. A quick characterization of materials for spintronics applications requires an essential piece of information: the spin polarization of currents in a specific direction. Our work is built around this central idea, which is similar to the mechanisms introduced in an earlier publication \cite{muller_spin_2009}. We perform ultrafast spin dynamics experiments using the time-resolved magneto-optical Kerr effect (tr-MOKE) to benchmark the spintronic properties in RuO$_2$. In the case of half metals, the spin polarization is determined via the demagnetization time. A fundamental principle prohibits spin scattering at the Fermi level for half metals, e.g., Fe$_3$O$_4$, CrO$_2$, and La$_{0.66}$Sr$_{0.33}$MnO$_3$, leading to an increased demagnetization time $\tau_M$ by two orders of magnitude, which is directly related to the degree of spin-polarization $P$. In altermagnets, we show that we can benchmark the spin-splitting and spin polarization of the bands in materials with spin-split electron bands by inducing an ultrafast coherent spin polarization, using circularly polarized laser pulses (Raman model). The Raman coherence time $\tau_{\textbf{R}}$ creates a temporal delay between the complex Kerr angle components, allowing to trace fingerprints of their altermagnetic properties.


Fig. \ref{fig:MaterialsOverview} explains the underlying principle of our approach by comparing the characteristics of three intensively investigated material classes for spintronics applications, ferromagnets, antiferromagnets, and topological insulators, with the typical properties of a d-wave altermagnet. With Figs. \ref{fig:MaterialsOverview} a) and b) pointing out the fundamental crystal symmetries and the spin alignments of the materials. Ferromagnets and antiferromagnets exhibit parallel or antiparallel spin orientations (see Fig. \ref{fig:MaterialsOverview} a)), respectively, with corresponding spin-split or non-spin-split Fermi surfaces (see Fig. \ref{fig:MaterialsOverview} b)). Topological insulators do not necessarily have assigned magnetic net moments, but due to spin-orbit interaction, electrons in surface layers form localized circular currents with a locked spin-momentum. This symmetry is broken along the surfaces, enabling direction-dependent spin currents \cite{fruchart_introduction_2013}. Altermagnets, such as RuO$_2$, possess spin-opposite sublattices with antiparallel alignment and therefore a compensated magnetic moment. The magnetization density of the magnetic atoms in the lattice is anisotropic, leading to time-reversal symmetry breaking and the appearance of the anisotropic spin-splitting (c.f. Fig. \ref{fig:MaterialsOverview} a),b)).

For all these four material classes, different transitions can be selectively induced by opposite photon helicities, as shown in Fig. \ref{fig:MaterialsOverview} c). In ferromagnets and antiferromagnets, these excitations are mainly $\vec{k}$ independent. However, for topological insulators and altermagnets, the excitation by circularly polarized photons strongly depends on the $\vec{k}$ direction. We investigate this band splitting by applying femtosecond-laser-pulse-driven excitation with opposite helicities, while probing the temporal evolution of both complex Kerr-angle components, the Kerr rotation $\Delta\theta_K$ and the Kerr ellipticity $\Delta\epsilon_K$. The induced transient spin-polarization proved to be coherent with the femtosecond laser pulse excitation. For altermagnets and topological insulators, the complex Kerr angle shows a specific time delay between the real part, $\Delta\theta_K$, and the imaginary part, $\Delta\epsilon_K$, for Raman times $\tau_R>0$ (c.f. Fig. \ref{fig:MaterialsOverview} d)). We dub this characteristic shape the "butterfly" structure. This signature is already established for topological insulators \cite{boschini_coherent_2015}, while this manuscript aims to discuss the butterfly structure in altermagnets. Although coherent induced effects were first identified and analyzed in ferromagnets by Bigot et al. \cite{bigot_coherent_2009}, generally demagnetization effects dominate the spin-dynamics processes in these materials. They suppress potential time delays between the complex Kerr-angle components, resulting in a linear relation between them \cite{walowski_energy_2008}.



One intensively explored candidate material that promises d-wave altermagnetic ordering is RuO$_2$. It would have the benefits, but also overcome the limitations of traditional antiferromagnetic spintronics \cite{wadley_electrical_2016}, as it exhibits spin-split bands $\sim 1200\,\mathrm{meV}$ and a critical temperature above room temperature ($\sim 400\,\mathrm{K}$) \cite{smejkal_emerging_2022}. But this material is not without controversy. On the one hand, Valenti, Mazin, and coworkers \cite{smolyanyuk_fragility_2024} calculated an extraordinary strain necessary to realize magnetic ordering in RuO$_2$. In addition, muon spin relaxation spectroscopy and neutron scattering experiments \cite{kesler_absence_2024, hiraishi_nonmagnetic_2024} as well as spintronic THz emitter application experiments \cite{plouff_revisiting_2025} point to the absence of magnetic order in bulk and ultra-thin films. This outcome is supported by several other research groups in various experiments \cite{liu_absence_2024, kiefer_crystal_2025, song_absence_2025, wang_absence_2026}. On the other hand, further research groups reported signatures of the anomalous Hall effect \cite{feng_anomalous_2022, tschirner_saturation_2023}, spin-polarized currents \cite{bose_tilted_2022, karube_observation_2022, bai_observation_2022, zhang_electrical_2025}, and spin-split bands in angle-resolved photoemission spectroscopy measurements \cite{fedchenko_observation_2024, lytvynenko_magnetic_2026} for the proposed d-wave altermagnet RuO$_2$. Those insights imply altermagnetically ordered moments, as their presence is necessary to measure the listed effects. More recent studies reveal that a closer look at the discussed phenomena is crucial due to their entanglement with other spintronic effects, e.g. the anisotropic (inverse) spin Hall effect in $\mathrm{RuO_2}$ \cite{wang_absence_2026, jechumtal_spin--charge-current_2025}. These contradictory experimental results highlight the balance act required to magnetically order $\mathrm{RuO_2}$, which is heavily influenced by sample parameters, e.g., strain, which varies depending on substrate choice, layer thickness, and growth conditions \cite{brahimi_confinement-induced_2025, meinert_meta-gga_2025, meinert_neel_2026}. 

\begin{figure}[!ht]
    \centering
    \includegraphics[width=0.75\linewidth]{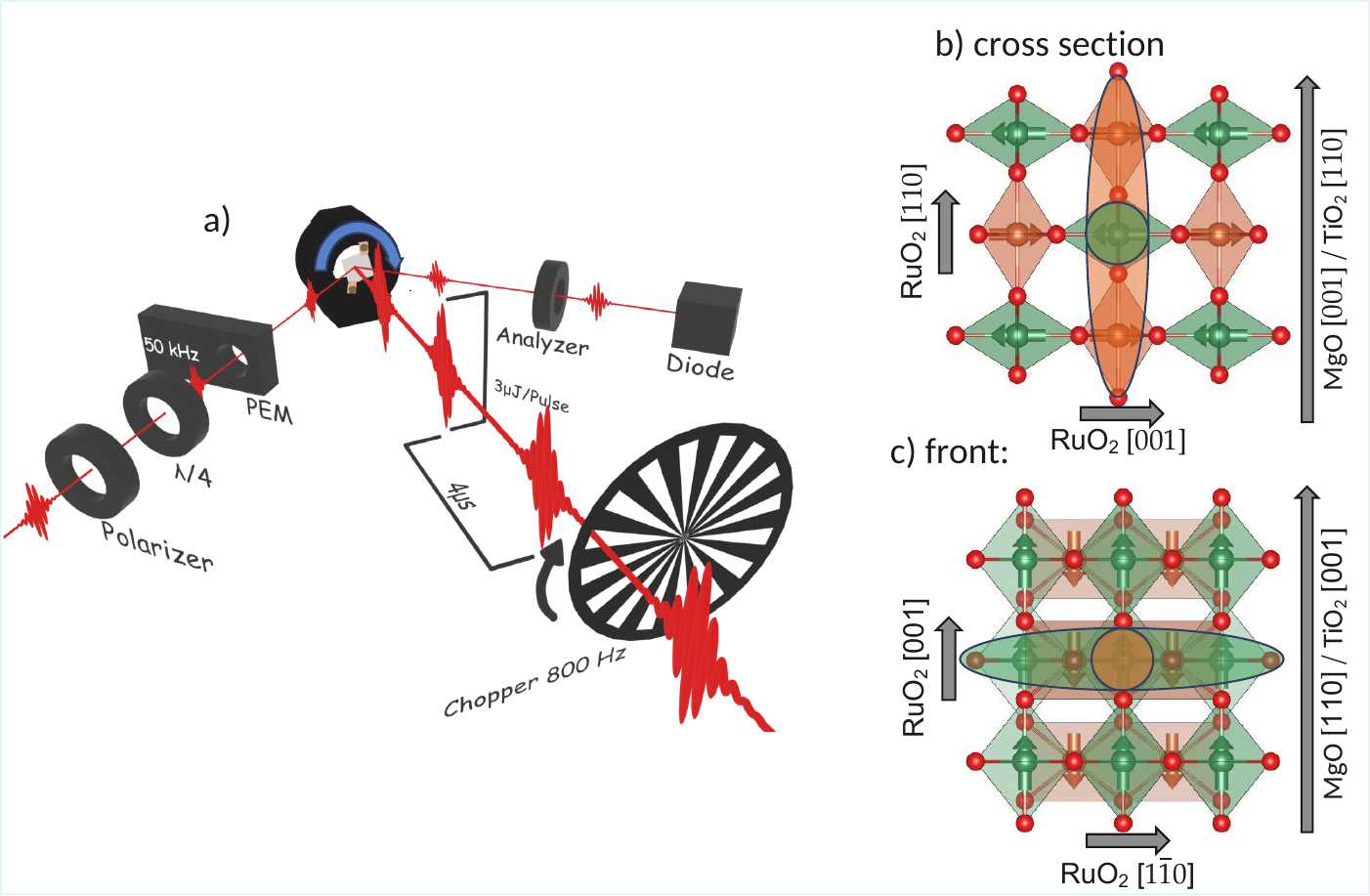 }
    \caption{The setup a) employs a double modulation time-resolved pump-probe MOKE scheme to extract dynamics on the femtosecond timescale. A photo-elastic modulator modulates the probe, and a mechanical chopper modulates the pump beam. The sample is rotated by the angle $\alpha$ indicated by the blue arrow. Crystal structure overview, cross-section b), and front view c). The grey arrows mark the crystal directions. The Néel vector points in the [001] direction. Oxygen atoms are colored red, Ru atoms in either sublattice are colored orange/green with their corresponding magnetic moment hinted by arrows. The oval and circular structures in orange and green indicate a cut through the Fermi surface.}
    \label{fig:Setup}
\end{figure}

Recent low-energy muon spin relaxation spectroscopy with depth-resolved sensitivity performed in transverse magnetic fields from $4\,\mathrm{K}$ to $290\,\mathrm{K}$ using muon implantation energies around $1\,\mathrm{keV}$ reveals magnetic signals originating from surface regions ($\sim10\,\mathrm{nm}$) in $30\,\mathrm{nm}$ thick films \cite{akashdeep_surface-localized_2026}. The authors find that these signals remain robust throughout various substrate and growfth conditions. These findings underscore the importance of reduced-dimensional contributions and motivate further investigations into the role of defects, strain, and stoichiometry for the magnetic properties of RuO$_2$, especially at its surface. Therefore, surface-sensitive experiments, including magneto-optical Kerr effect (MOKE) measurements, are the methods of choice to investigate the electron band spin-splitting of these magnetic surface states. We use a pump-probe scheme depicted in Fig. \ref{fig:Setup} a) to investigate spin dynamics.

Altermagnetic signatures appear in the change of both Kerr angle components as functions of time delay $\Delta\tau$ after excitation by a pump pulse. Fig. \ref{fig:MaterialsOverview} d) shows the expected relation for the change between both Kerr angle components, the Kerr rotation, and the Kerr ellipticity on a femtosecond time scale for the underlying band structures of the magnetic materials introduced above. In this experiment, ferromagnets demagnetize ultrafast on the sub-picosecond timescale, followed by a slow recovery lasting tens of picoseconds regardless of photon helicity. This ultrafast demagnetization process leads to a change in $\Delta\theta_K$, while $\Delta\epsilon_K$ reacts to the shift in the electron population. Both changes occur simultaneously. This mechanism holds for antiferromagnets, where THz spin precession dominates the signal. In topological insulators, circularly polarized photons selectively excite electrons with the corresponding $\vec{k}$ and its associated spin. This process, combined with the characteristic decay, manifests itself in the experiment as a time delay between $\Delta\theta_K$ and $\Delta\epsilon_K$, yielding a butterfly structure in the complex plane spanned by both components of the Kerr angle. Finally, altermagnets exhibit $\vec{k}$ direction-dependent spin states, acting similarly to spin-momentum-locked states in topological insulators. Those states initiate the same characteristic patterns in the complex plane as observed in topological insulators, the butterfly patterns. Thus, different shapes of the components of the real and imaginary Kerr angle are created via the Kramers-Kronig relation \cite{boschini_coherent_2015}.

We investigate $d\approx20\,\mathrm{nm}$ thick RuO$_2$ films with a (110) orientation and a $3\,\mathrm{nm}$ Pt capping, grown on (110)-oriented TiO$_2$ or (001)-oriented MgO substrates. In this geometry, we rotate the samples and thus the opposite-spin sublattices around the [110] direction axis by an angle $\alpha$. Typically, the Neél vector points in the RuO$_2$ [001] direction. Therefore, in our geometry, the spin orientation remains in the sample plane, together with the spin-split Fermi surface from one sublattice, leaving the spin-split Fermi surface of the other sublattices directed out-of-plane (see Fig. \ref{fig:Setup} b),c)). Looking from the [110] direction axis, a spin splitting occurs in the $\Gamma$ point to $M$ point direction, but not in the $\Gamma$ point to $Z$ point direction \cite{osumi_spin-degenerate_2026}. We excite the dynamics by employing linear ($\pi$) (electric field vector parallel to the plane of incidence or p-polarization) and circularly ($\sigma^{\pm}$) polarized pump pulses to elaborate differences in the momentum transfer from the photons. For this, we record the variations in the Kerr rotation $\Delta\theta_K$ and the Kerr ellipticity $\Delta\epsilon_K$ as functions of the time-delay $\Delta\tau$ after excitation by pump pulses. We record these dynamics for a complete sample rotation around the surface normal axis in $\alpha=15^{\circ}$ steps with respect to the set pump-pulse polarization. An 800 nm (1.55 eV) Ti:Sapphire laser oscillator with a regenerative amplifier at a repetition rate of 250 kHz delivers the pulses with a FWHM $t_{\mathrm{pulse}}$ duration of $40\,\mathrm{fs}$. Figure \ref{fig:Setup} a) shows a simplified sketch of the measurement setup. We employ a double modulation detection scheme. A photo-elastic crystal modulates the probe beam polarization for MOKE signal detection, and an optical chopper in the pump beam path enhances the temporal MOKE change signal-to-noise ratio after excitation.

\section{Discussion}
\begin{figure}[!ht]
    \centering
    \includegraphics[width=0.95\linewidth]{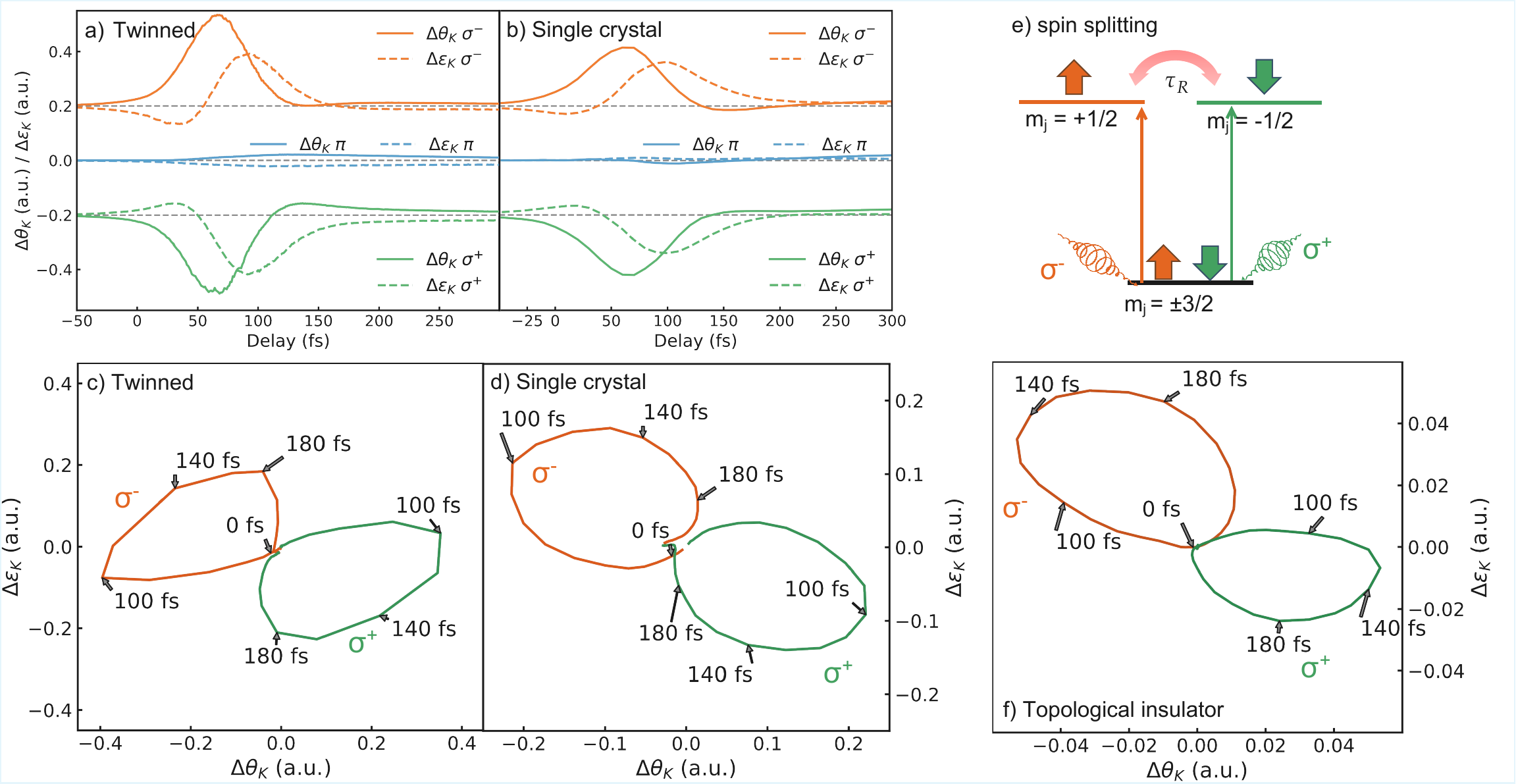}
    \caption{Time-resolved complex Kerr angle components data for a) RuO$_2$ on MgO, and b) RuO$_2$ on TiO$_2$ during the first $300\,\mathrm{fs}$ after excitation for all three excitation polarizations. The $\sigma^{\pm}$ spectra are shifted for clarity; the gray dashed lines represent the original zero line. The circular signal is on average $100\,\mathrm{fs}$ faster and 10 times stronger than the linear response. Additionally, a second peak arises. Dynamics data transferred to the complex plane (c) and d)), depicting a butterfly structure, which develops during the femtosecond scattering process between the transient spin polarized states e). Also observed in topological insulators f).}
    \label{fig:RuO2 Dynamics and complex plane}
\end{figure}


Spin dynamics in RuO$_2$, generated by exciting the electron system with linearly polarized pulses, are discussed in reference \cite{weber_all_2024}. Therefore, we focus on the dynamics generated using circularly polarized laser pulses. The temporal evolution of $\Delta\theta_K$ and $\Delta\epsilon_K$ up to $\Delta\tau=300\,\mathrm{fs}$  for a representative sample orientation of $\alpha=60^{\circ}$ shows the same features for both RuO$_2$ samples, as Figs. \ref{fig:RuO2 Dynamics and complex plane} a) (on MgO) and b) (on TiO$_2$) depict. A linear pump ($\pi$, blue lines) generates a small demagnetization peak, followed by a remagnetization process on the picosecond (ps) timescale. Both signals $\Delta\theta_K$ and $\Delta\epsilon_K$ are inverted.
The right-circular $\sigma^-$ (orange) and left-circular $\sigma^+$ (green) pumps generate a temporal evolution, where $\Delta\theta_K$ exhibits a pronounced peak (positive for $\sigma^-$ and negative $\sigma^+$) followed by a distinct dip. For $\Delta\epsilon_K$, the sequence and sign between the peak and the dip reverse. Both spectra are shifted for the sake of clarity, with the gray dashed line representing the original zero line. Their trends are not completely independent. They combine the slope of the Gaussian peak with the width of the laser excitation and its derivative, with different amplitudes and signs, in sequence, suggesting a relative time delay between the two components. Nevertheless, the pronounced symmetry of the Kerr response under $\sigma^-$ polarized excitation, together with the negligible reflectivity contribution from the MgO substrate, points to a magnetic origin. Both signals reverse sign upon inversion of the pump helicity. 


This seemingly complex behavior of a coherently laser-driven ultrafast spin polarization bears a quite simple explanation. Its physical origin is related to selection rules and spin-orbit interaction, and thus it is a signature of transient spin-polarized states. This behavior was observed in semiconductors by \textit{A. Kimel et. al.} \cite{kimel_room-temperature_2001} and was explained within a simple theoretical model. They investigated the band gaps in intrinsic and n-doped GaAs using time-dependent Kerr spectroscopy and, based on their findings, developed a model that reveals the microscopic processes triggered by the optical pumping on ultrafast timescales. Their calculation of the full complex Kerr angle assumes a simple band model with four energy levels that accounts for spin-split bands arising from spin-orbit interaction in both the ground and the excited states (compare Fig. \ref{fig:RuO2 Dynamics and complex plane} e)). In such a model, selective pumping processes, assigned by the photon helicity, populate only one of the two possible excited states, thereby creating a transient spin polarization that decays on a characteristic timescale, the Raman coherence time $\tau_R$. The typical pattern observed in the complex Kerr rotation is interpreted as proof of the existence of spin-polarized bands. Scattering processes, during which excited electrons distribute between different m$_\mathrm{J}$ states, define $\tau_R$. Several investigations tested the model on CuO$_4$ based antiferromagnetic Mott insulators ($\tau_R\approx30-50\,\mathrm{fs}$ \cite{pavlov_ultrafast_2007}), the multiferroic material YMnO$_3$ ($\tau_R\approx10\,\mathrm{fs}$ \cite{pohl_ultrafast_2013}). We have already adapted this method to successfully characterize spin states in the topological insulator $\mathrm{(BiSb)_2Te_3}$ ($\tau_R\approx9-15\,\mathrm{fs}$ \cite{boschini_coherent_2015}). Using that band model with spin-split bands, an analytic function for the complex Kerr rotation can be derived:
\begin{equation}
    \theta_K+i\epsilon_K = A \exp\left(-\frac{\Delta\tau^2}{w^2}\right)+B \exp\left(\frac{w^2}{4\tau_R^2}-\frac{\Delta\tau}{w}\right)\left(1-\mathrm{erf}\left(\frac{w}{2\tau_R}-\frac{\Delta\tau}{w}\right)\right),
    \label{eq:complex Kerr angle}
\end{equation}
where $w=t_{\mathrm{pulse}}/\sqrt{2\ln(2)}$ is the pulse duration, and $A$, $B$ are the complex amplitudes arising from the dielectric susceptibility tensor \cite{pavlov_ultrafast_2007, pohl_ultrafast_2013}. The first term represents the direct response to laser excitation, indicating a selective pumping process, whereas the second term describes Raman scattering. Fitting eq. \ref{eq:complex Kerr angle} to the RuO$_2$ Kerr dynamics data reveals a $\tau_R\approx 4.5-7\,\mathrm{fs}$, which is shorter than for the other material systems, but in the same order of magnitude. Although the Kerr signal changes, $\tau_R$ is independent of the excitation angle. Therefore, we see strong evidence for transient spin-polarized states in RuO$_2$, optically induced and equilibrating on the femtosecond timescale. In previous studies, the authors applied this model to investigate material systems with spin-split bands, using it as a probe to verify their presence.

In Figs. \ref{fig:RuO2 Dynamics and complex plane} c) and d), the Kerr angle components $\Delta\theta_K$ and $\Delta\epsilon_K$ are transferred to the complex plane for both samples, RuO$_2$ grown on MgO and TiO$_2$, respectively. 

Finally, Fig. \ref{fig:RuO2 Dynamics and complex plane} e) illustrates the impact of the quantum mechanical angular momentum conservation law for a simplified RuO$_2$ band structure in the spin-split $\Gamma-M$ direction, employing a simple energy level scheme. The selection rules prohibit specific electron transitions, e.g., excitations from the ground state $m_j=3/2$ by $\sigma^+$ photons. This prohibition increases the transition probabilities from $m_j\neq 3/2$ states relative to transitions induced by $\pi$-polarized photons. A similar effect is discussed in reference \cite{boschini_coherent_2015}, where ultrafast spin dynamics in topological insulators is explained, based on a two-level band model. Both material systems have one thing in common: increased photon densities, as are concentrated in femtosecond laser pulses, enhance higher-order excitation probabilities, leading to demagnetization effects. For first-order excitations, however, the photons induce coherent, spin-polarized states that scatter on a femtosecond timescale. The crucial difference between the two systems is that RuO$_2$ is not spin-split along the $\Gamma-Z$ direction, and we do not expect a distinct pumping behavior with respect to opposite helicities. Additionally, in topological insulators, scattering processes are always associated with a reversal of the $\vec{k}$ direction upon a spin flip, due to spin-momentum locking. In RuO$_2$, the symmetry of the electron band associates a spin flip with a direction change of $\vec{k}$ by $90^{\circ}$. Both processes result in a similar relation between $\Delta\theta_K$ and $\Delta\epsilon_K$, forming butterfly structures, see Figs. \ref{fig:RuO2 Dynamics and complex plane} c), d) for RuO$_2$ and f) for a topological insulator (adapted from \cite{boschini_coherent_2015}).

Density functional theory calculations predict two crystalline directions for spin-split bands in rutile RuO$_2$ single crystals that are perpendicular to each other and occupied by electrons with opposite spin \cite{osumi_spin-degenerate_2026}. These spin-split bands in d-orbitals form along the long tetragonal axes, as illustrated in Figs. \ref{fig:Setup} b) and c). With respect to the (110) plane, one of those spin-split bands is oriented in-plane, while the second band lies perpendicular to the (110) plane, out-of-plane. We expect an angle-dependent spin-dynamics signal from the in-plane-oriented band, while the signal from the out-of-plane band should be angle-independent. It is reasonable to assume $\alpha=180^{\circ}$ periodicity of the spin polarization, mirrored in $\Delta\theta_K$, and therefore a strong response along the in-plane, spin-split direction ($135^{\circ}$), and a vanishing signal along the perpendicularly oriented in-plane axis ($45^\circ$) where no splitting occurs. The out-of-plane spin-split bands retain their orientation relative to the incoming photons during sample rotation. This induces a constant signal, either positive or negative, depending on the light's helicity. The superposition of both signals (in-plane and out-of-plane) generates an oscillating pattern $\propto \sin^2(x)$ with a non-vanishing constant sign amplitude.

\begin{figure}[!ht]
    \centering
    \includegraphics[width=0.95\linewidth]{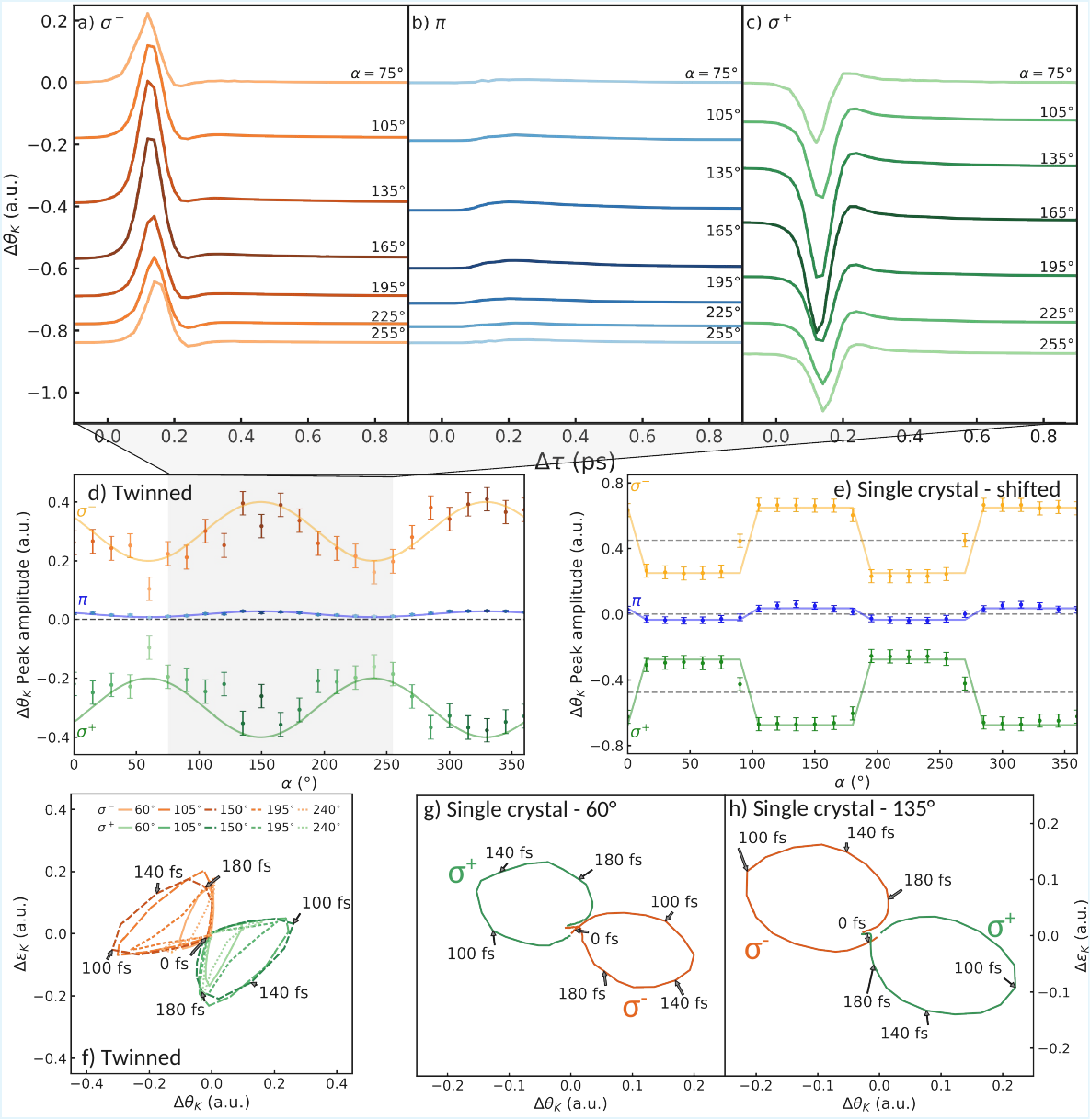}
    \caption{RuO$_2$ on MgO: $\Delta\theta_K$ spectra for an a) left circular, b) linear, and c) right circular polarized excitation during the first 1 ps after excitation for selected sample rotation angles. d) and e), the first peak extracted from the corresponding dynamic curves for RuO$_2$ on MgO (twinned) and TiO$_2$ (single crystal), respectively, plotted as a function of sample rotation angle $\alpha$. The data show a clear twofold symmetry for a full sample rotation. The single-crystal data are shifted for clarity, with dashed lines denoting the zero-crossing. In the complex plane, f) twinned and g), h) single crystal, the butterfly structure is shown for both samples. In g) and h), the "switching" behavior shows in the sign change between the wings from $\alpha=60^{\circ}$ and $\alpha=135^{\circ}$.}
    \label{fig:RuO2 MgO Overview Plots}
\end{figure}

Given these considerations, we conduct a systematic angle-dependent study of the spin dynamics of our samples. We rotate the samples around the [110] direction axis by angle $\alpha$, as described in Fig. \ref{fig:Setup} a) (blue arrow) and b) - c). Figs. \ref{fig:RuO2 MgO Overview Plots} a)-c) present the time-resolved Kerr rotation dynamics ($\Delta\theta_K$) for selected sample rotation angles $\alpha$ from $75^{\circ}$ to $255^{\circ}$, up to $\Delta\tau=1\,\mathrm{ps}$ for the RuO$_2$ thin film grown on MgO, employing right circularly $\sigma^{-}$, linearly $\pi$ and left circularly $\sigma^{+}$ polarized laser pulses. The Kerr response is qualitatively similar, regardless of excitation polarization. The magneto-optical response peaks upon the arrival of the pump pulse and relaxes on a picosecond timescale. However, two aspects stand out. First, ($\sigma^{\pm}$) pulses generate tenfold larger Kerr signals, and these emerge $\sim 100\,\mathrm{fs}$ earlier than those generated by ($\pi$) pulses. Second, $\sigma^{+}$ and $\sigma^{-}$ pulses generate Kerr dynamics of opposite sign at ultrashort delays ($\Delta\tau<100\,\mathrm{fs}$).
The absolute amplitude of the transient $\Delta\theta_K$ peak increases as $\alpha$ increases from $75^{\circ}$ to $135^{\circ}$, reaches a plateau around $\alpha\approx150^{\circ}$, and subsequently decreases to a minimum at $\alpha\approx 255^{\circ}$.

Those peak amplitudes for a full rotation are extracted in Fig. \ref{fig:RuO2 MgO Overview Plots} d) for RuO$_2$ grown on the MgO substrate. They exhibit identical periodic angular dependencies with pronounced twofold symmetry across all pump polarizations. For a single-crystalline rutile structure, we expect a pattern $\sim\sin^2(\alpha)$ in the extracted peak amplitudes as a function of the rotation angle $\alpha$. However, additional symmetry aspects shift this pattern from the zero line. First, the lattice mismatch between the cubic MgO substrate and the tetragonal RuO$_2$ causes a crystal twinning, forcing the deposited material into two crystal sub-systems rotated by 90$^{\circ}$. Additionally, they do not develop a well-defined Néel vector direction. From the perspective of the substrate, the RuO$_2$ crystals grow rotated by $\pm45^\circ$ against the [100] MgO direction, allowing two opposite Néel vector directions for each sub-crystal system. This allows for four possible magnetic ordering directions in $20-30\,\mathrm{nm}$ long, $5-10\,\mathrm{nm}$ wide, spatially separated, grain-like crystals with a 1:1 distribution. The probe beam spot covers an area $A_{\mathrm{probe}}\approx350\,\mathrm{\mu m^2}$, which is much smaller than the pump spot size $A_{\mathrm{pump}}\approx11000\,\mathrm{\mu m^2}$. That means we excite and probe thousands of grains, recording a superposition of $n$ Kerr signals $\Delta\theta_K^S,\,\Delta\epsilon_K^S \propto \sum_{n} a_n\cdot\sin^2(\alpha-\alpha_\mathrm{offset,n})$, generated at crystal evenly distributed crystal sub-systems. For such a distribution, $a_n = 1$ or $a_n = -1$ represent the two possible Néel vector directions and magnitudes, and $\alpha_\mathrm{offset,n}=\pm 45^\circ$ represents the two crystal twins. Such a configuration would result in a signal $\Delta\theta_K^S,\,\Delta\epsilon_K^S$ independent of the sample rotation, which we do not measure. 

Asymmetries in the sub-crystal system do not account for the uneven distribution of the Néel vectors, because X-ray diffraction crystallography reveals an almost perfect parity between the crystal twin distributions, with a very small variation $< 1\%$. However, small magnetic fields, always present during crystal growth, can bias the direction of the Néel vector. Accordingly, one of the four possible Néel vectors may dominate the signal and lead to an angle dependence. This results in a non-vanishing amplitude, following the $\sim\sin^2(\alpha)$ pattern shifted from zero. This behavior resembles the inflating and deflating "butterfly" pattern in the complex plane, shown in Fig. \ref{fig:RuO2 MgO Overview Plots} f). Their shape demonstrates an angle dependence for $\Delta\theta_K$, indicating the strength of the induced transient spin-polarized state. At the same time, $\Delta\epsilon_K$ remains almost constant throughout the sample rotation, implying that electrons are pumped into bands with the same orbital momentum independent of the excitation vector.

We explain the small Kerr signal amplitudes and missing second oscillation peak measured for $\pi$ polarized excitation on both substrates by considering linear polarization as a superposition of $\sigma^+$ and $\sigma^-$ polarizations. In that case, all electron bands are excited equally, limiting a distinct spin-separated response. We attribute the remaining magnetic response signal to long-lived spin-polarization, as discussed in reference \cite{weber_all_2024}, which seems also present in the (110) crystal structure, in which one sublattice band structure is oriented in-plane, and the other is oriented out-of-plane.

Noticeably, we observe a sign change in the RuO$_2$ sample grown on TiO$_2$. Fig.\ref{fig:RuO2 MgO Overview Plots} e) shows the extracted Kerr signal peak amplitudes plotted as a function of the sample rotation angle $\alpha$ for a full rotation, analogously to the data in Fig. \ref{fig:RuO2 MgO Overview Plots} d) (data are shifted for clarity, the gray dashed lines represent zero crossing, orange/green are the $\sigma^-/\sigma^+$ excitation data). The data follow a pattern similar to that for RuO$_2$ grown on MgO, but instead of a trend $\sim\sin^2(\alpha)$, a much sharper step-like switching occurs. In addition, each sample rotation by $\alpha=90^{\circ}$ induces a polarization sign change, see Figs. \ref{fig:RuO2 MgO Overview Plots} g) and h).



In conclusion, our data confirm that ultrashort circularly polarized laser pulses induce short-lived magnetically polarized states in RuO$_2$, and we can control their polarity by reversing the helicity of the pump pulse and changing the crystal orientation.
We demonstrate the existence of transient spin-polarized states by analyzing the characteristic relation between both Kerr-angle components measured as a dynamic response to excitation by ultrashort laser pulses, resulting in a "butterfly" structure. This analysis enables us to establish a link between the findings for RuO$_2$ crystals and materials with spin-split bands, e.g., topological insulators. The extracted Raman scattering time $\tau_R<10\,\mathrm{fs}$ after which the spin polarization decays denotes the lifetime of the spin-separated states. The strength of polarization changes with the crystalline direction of the sample, according to the symmetries of the proposed spin-split bands of the crystal \cite{osumi_spin-degenerate_2026}. Consequently, this new non-contact method can be used to characterize altermagnets fast and without destroying the sample properties. Excitation by linearly polarized photons does not occupy spin-separated states above the Fermi level selectively, resulting in a Kerr signal that is one order of magnitude smaller. Finally, excitation by circularly polarized photons enables us to generate controlled, coherent transient spin-polarized states driven by polarized light pulses on an ultrafast timescale.

\bibliographystyle{unsrt}
\bibliography{RuO2Paper.bib}

\section{Supplementary}


\subsection{Sample preparation}

\begin{figure}[!h]
    \centering
    \includegraphics[width=0.9\linewidth]{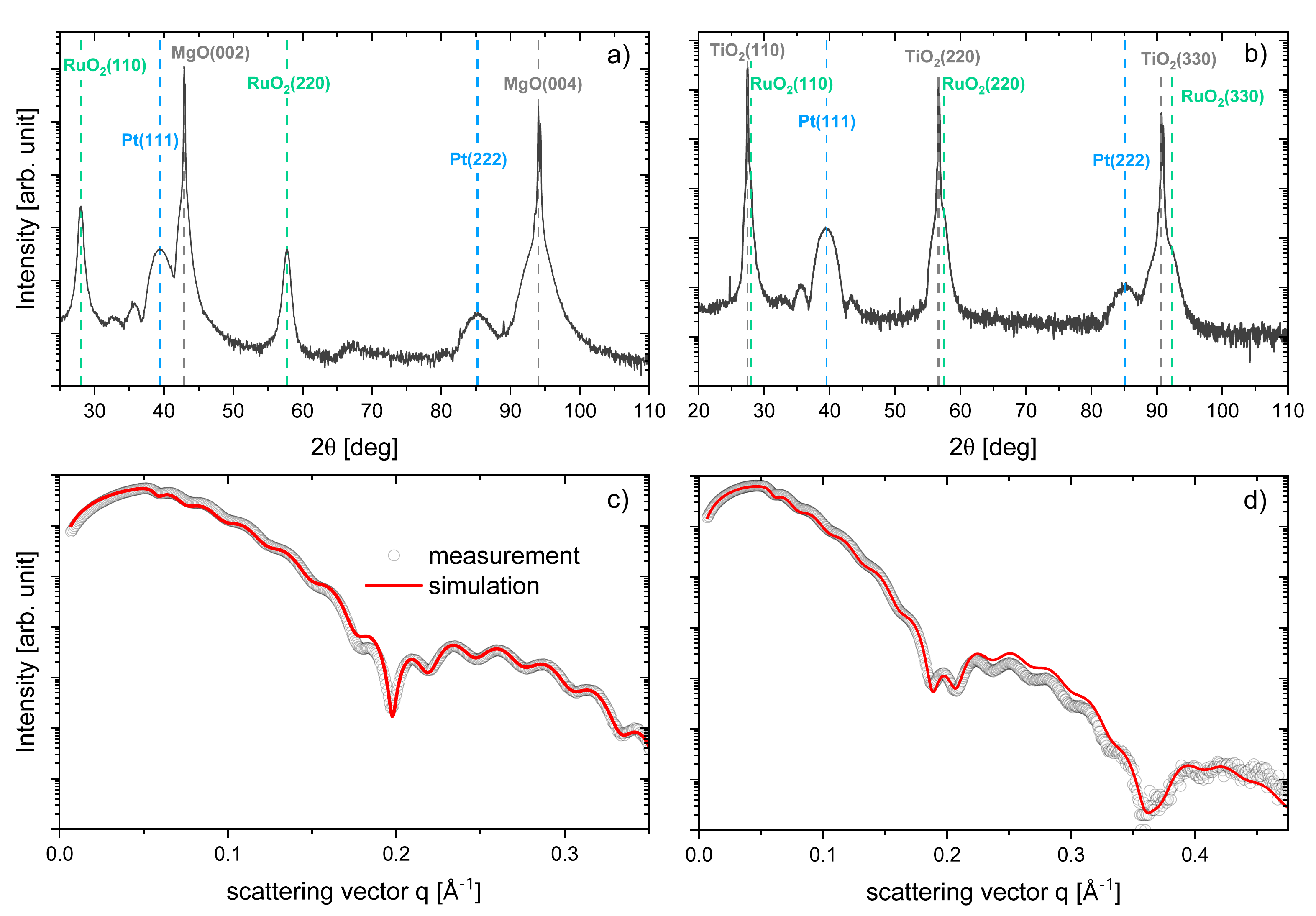}
    \caption{a) and b) X-ray diffraction spectra for both samples grown on MgO and TiO$_2$, respectively. c) and d) X-ray reflectivity spectra for both samples grown on MgO and TiO$_2$, respectively.}
    \label{fig:Sample_characterization}
\end{figure}

The RuO$_2$ layers are grown by reactive RF-magnetron sputtering using an Ar:O$_2$ ratio of 4:1 in a UHV chamber with a base pressure below $1\cdot 10{-9}$\,mbar. The RuO$_2$ layer is deposited onto MgO(001) and TiO$_2$(110) substrates at substrate temperatures $\approx230^{\circ}\mathrm{C}$. The Pt capping is deposited in-situ at room temperature by DC-magnetron sputtering. The samples are then characterized by X-ray diffraction (XRD) and X-ray reflectivity (XRR) in a PANalytical X’Pert Pro MPD PW3040-60 diffractometer using Cu K$\alpha $ radiation. RuO$_2$(ll0) peaks appear after deposition on both substrates, as illustrated in Figs. \ref{fig:Sample_characterization} a) and b). In both cases, the Pt-capping is (111) textured with pronounced Laue Oscillations. XRR measurements for samples grown on MgO(001) and TiO$_2$(110) are shown in Figs. \ref{fig:Sample_characterization} c) and d), respectively. Layer thicknesses $d$ and roughnesses $\sigma$ are determined using the open-source software GenX \cite{Bjoerck2007} and are presented in the Tab. \ref{tab:Sample_parameters}.

\begin{table}[!ht]
\caption{Sample parameters.}
\center
\begin{tabular}{|ccc|lll|}
\hline
\multicolumn{3}{|c|}{sample grown on MgO(001)} & \multicolumn{3}{l|}{sample grown on TiO$_2$(110)} \\ \hline

layer     & $d$ {[}nm{]}  & $\sigma $ {[}nm{]}  & layer      & $d$ {[}nm{]}   & $\sigma $ {[}nm{]}   \\ \hline

Pt        & 3.5         & 0.6                 & Pt         & 3.6          & 0.6                  \\

RuO$_2$   & 19.8        & 0.5                 & RuO$_2$    & 18.0         & 0.5                  \\

MgO       & -           & 0.5                 & MgO        & -            & 0.6          \\      
\hline
\end{tabular}
\label{tab:Sample_parameters}
\end{table}

\end{document}